# Measurement of energy differential spectrum of cosmic-ray muons below 400 MeV

Hikaru Sato[1], Tadahiro Kin[1], Andrea Giammanco[2]

[1] Interdisciplinary Graduate School of Engineering Sciences, Kyushu University, Fukuoka, Japan
[2] Centre for Cosmology, Particle Physics and Phenomenology, Université catholique de Louvain, Louvain-la-Neuve, Belgium

E-mail: sato.hikaru.582@s.kyushu-u.ac.jp

**Abstract**

Recent applications of cosmic-ray muons require accurate modeling of their flux at low-energy. However, no measurement has been reported below 400 MeV. Therefore, we developed a full-absorption muon energy spectrometer to obtain energy differential flux below 400 MeV. Because our main detector can measure muon energies below 75 MeV, an energy degradation method is adopted (using 5- and 20-cm thick lead blocks) to shift the sensitive energy range. Three measurements were performed (in the normal mode and the two energy degrading modes) for 10 and 11 days each. The measurement results were compared with an analytical cosmic-ray model, PARMA—the particle and heavy ion transport code system-based analytical radiation model in the atmosphere—and we found that the model can precisely predict the lower energy part of the flux.

## 1. Introduction

The double-differential energy spectra of terrestrial cosmic-ray (CR) muons are essential in numerous applications, such as muography and soft error rate estimation. First, we consider the requirements in the field of CR muon radiography or simply called muography [1]. Muography is a nondestructive inspection method for the internal structure of large objects, and it is based on the absorption of muons that depends on the materials amount. Its history started with a search for hidden chambers in pyramids by Alvarez et al. [2]. Recently, muography has attracted research attention in various fields. For example, the internal structure of volcanoes has been investigated by near-horizontally arriving CR muons [3–9]. At present, the application of muography is expanding to smaller objects, such as dams and bridges; moreover, portable detectors have been developed for multipurpose applications [10–13]. For example, Chaiwangkhot et al. [10] developed a portable muography detector to inspect infrastructure buildings. For the inspection of small objects, within a few to a few tens of meters, an essential role is played by the absorption of muons of energy below 1 GeV. Second, we consider the requirements of CR muon spectra in the soft error field. Cosmic rays are one of the causes of soft errors in microelectronics. Cosmic-ray neutrons contribute considerable to soft errors, and they have recently become unignorable. The cause is that state-of-the-art devices need a low operating voltage to achieve low-power consumption, and this trend implies that the minimum charge required to change a memory state to a different value is getting minute, so even the modest energy deposited by a CR muon can cause a soft error. The precise estimation of soft errors due to CR muons requires a good model of the low-energy CR muon component that can deposit enormous energy to a small sensitive region of a device and produce a higher charge than the soft error minimum charge [14,15].

Therefore, in both applications, energy spectra in the low-energy region are essential to estimate the attenuation ratio in the smaller target muography and to predict deposite energy for accurate soft error rate. However, only a few measurements have been reported below 1 GeV, and there are completely no measured data below 400 MeV. Therefore, low-energy CR muon spectra are predicted using analytical models, such as the particle and heavy ion transport code system-based analytical radiation model in the atmosphere (PARMA) [16,17] and the cosmic-ray shower library. These models have sufficient performance above 1 GeV, but the low-energy reproducibility has not been validated due to a lack of experimental data. To address this issue, we developed a full-absorption muon energy spectrometer (FAMES) to obtain energy differential spectra below 400 MeV.

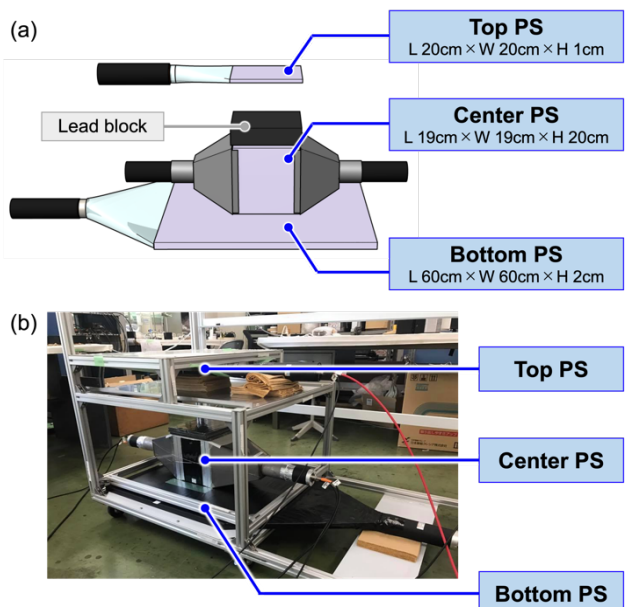

**Figure 1.** (a) Schematic view and (b) photograph of FAMES. FAMES comprises three PSs to measure the kinetic energy of low-energy muons while removing background events.

## 2. FAMES

### 2.1 Features

#### Geometrical setup
We developed FAMES to measure the low-energy spectra of terrestrial CR muons. As shown in Figure 1, FAMES comprises three plastic scintillators (PSs): Top PS (L 20 cm × W 20 cm × H 1 cm), Center PS (L 19 cm × W 19 cm × H 20 cm), and Bottom PS (L 60 cm × W 60 cm × H 2 cm). The distance between Top and Center PSs was set to 25 cm to detect zenith angles between 0° and 40°.

#### Center PS
The main detector is a cubic PS located at the center of FAMES, Center PS, which aims to absorb all kinetic energy of the low-energy CR muon.

#### Top PS
PSs are sensitive to CR electrons and positrons as well as environmental gamma and beta rays. Top PS is placed above Center PS for coincidence detection to eliminate the environmental rays. The distance between the two PSs determines the detection acceptance of FAMES. However, CR electrons and positrons have sufficient energy to generate signals in both PSs, and they constitute approximately 20% of the counting rate when taking data in open-sky conditions. These particles have a few hundreds of MeV in kinetic energy, which overlap with the low-energy CR muon spectrum. Thus, FAMES employs the ΔE–E method for particle identification: Top PS acts as ΔE (thin) detector, whereas Center PS measures the total E of the stopped particles. In the two-dimensional (2D) histogram of ΔE versus E, CR electrons and positrons can be distinguished from muon events because their stopping power differs, attributable to the mass difference.

In addition, even under the coincidence condition of Top and Center PSs, events by random noise coincidence are significant for small analog-to-digital converter (ADC) channel numbers, for both Top and Center PSs.

#### Bottom PS
Bottom PS is placed at the bottom side of Center PS to detect muons that pass the coincidence condition and can exit from the side surface of Center PS. Bottom PS is designed to cover the entire solid angle of the Top and Center PS coincidence regions so that all escaping particles can be detected by Bottom PS.

If Bottom PS fires a signal in coincidence with Top and Center PSs, it is interpreted as a high-energy muon. Therefore, the signal is used as an anticoincidence (veto) condition to reject the penetrating muons. In other words, all recorded events are considered full-absorption events.

### 2.2 Dynamic energy range

The dynamic energy range of FAMES is determined by the range of muons in Center PS. The dominant interaction of muons with plastic is an ionizing process, and Figure 2 shows muon stopping power in plastic [16]. A continuous slowing down approximation (CSDA) range can be derived by integrating the stopping power inverse up to the initial muon kinetic energy. The kinetic energy of muon that corresponds to a CSDA range of 20 cm (the height of Center PS) is 75 MeV, which is the minimum energy to penetrate Center PS from its bottom surface. The maximum energy of a full-absorption muon, 83 MeV, corresponds to muons that penetrate a corner of Top PS and the farthermost corner of Center PS.

We call "normal mode" the measurement described so far, and "degrading mode" a method to measure higher energy muons. In the latter mode, lead blocks are inserted between Top and Center PSs to degrade higher energy muons down to an energy range that can be detected by Center PS. The maximum load is mechanically determined to be L 20 cm × W 20 cm × H 20 cm, which allows measurements up to 400-MeV muons. The degrading mode measurements cannot determine total energy by Center PS directly because a part of the energy is lost in the absorber. In addition, the absorber strongly induces random, Coulomb scattering of muons and changes the track length in Center PS. This contribution is corrected by solving an inversion problem (see Section 4.2).

The lowest energy cutoff of FAMES was 20 MeV. Below this value, the electrical noise of the data acquisition (DAQ) system prevents precise identification in the ΔE-E plot by overlapping with the CR events.

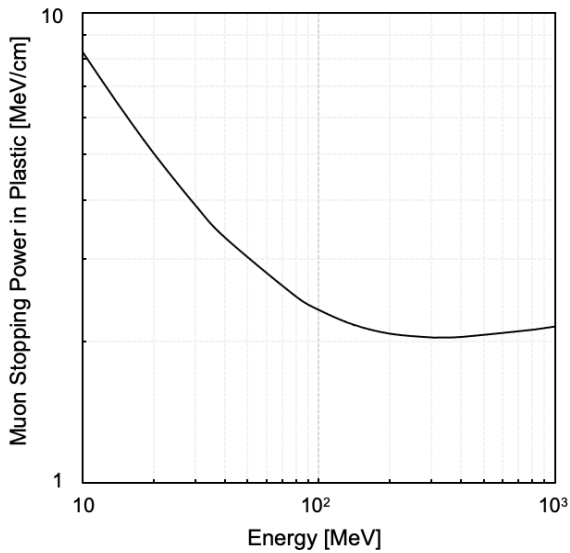

**Figure 2.** Stopping power for muons in plastic [16].

### 2.3 Suppression of decay event pile-up

Negative and positive muons decay to electrons and positrons, respectively, with 2.2 μs of average life when they are at rest. These decaying particles have a maximum kinetic energy of approximately 53 MeV [17] and deposit almost all their energy in Center PS. The pile-up event, the coincidental measurement of energy released by both the muon and its decay product as a single event, in which case the full pile-up charge is converted by the ADC in the DAQ system of FAMES, overestimating the kinetic energy of the incoming muon. This can be mitigated by generating a narrow timing gate signal in the DAQ system for the charge integral time of the ADC. The gate width is set to 100 ns to remove 95% of the pile-up events.

## 3. Data analysis methods

### 3.1 PHITS simulation

A Monte Carlo simulation code, PHITS, was employed for the following analysis. PHITS implements various models to simulate the ionizing process, which is dominant in determining the detector response. We chose the atomic interaction with matter model [18] for the stopping power calculation, the Landau–Vavilov distribution for energy struggling, and Lynch's formula based on the Moliere theory [19] for Coulomb scattering. In the energy calibration part, the spectral shape was determined using PARMA.

### 3.2 Energy calibration

The spectrum for Center PS has clear cutoff energy by the veto detection of Bottom PS at 75 MeV (Figure 3), as mentioned in Subsection 2.2. We used this cutoff and the ADC pedestal point to derive a linear calibration.

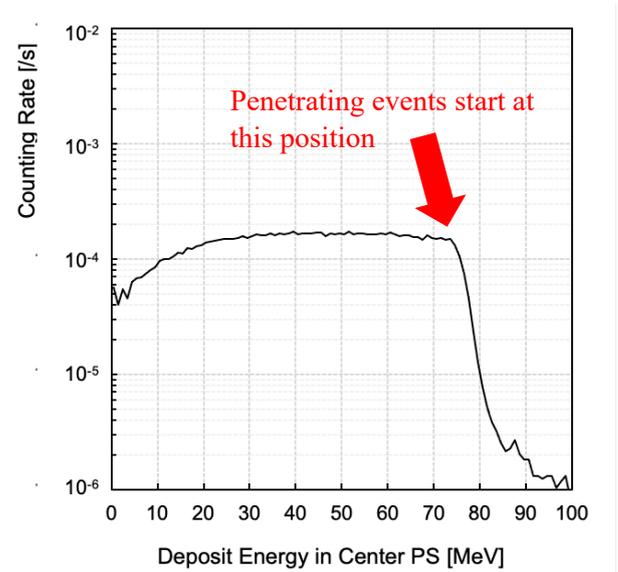

**Figure 3.** Counting rate as a function of energy deposited in Center PS. The rate drops sharply as muons begin to penetrate above 75 MeV.

Moreover, the Top PS simulation spectrum shows a clear Landau peak (Figure 4). This peak is created by the minimum ionized particles from the vertical direction. Using the Landau peak and ADC pedestal, a linear calibration function was derived.

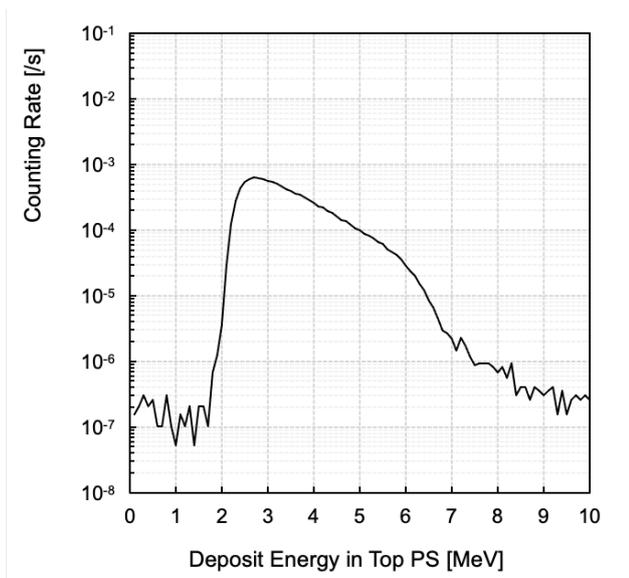

**Figure 4.** Counting rate as a function of energy deposited in Top PS. The minimum ionization peak is approximately 2.5 MeV.

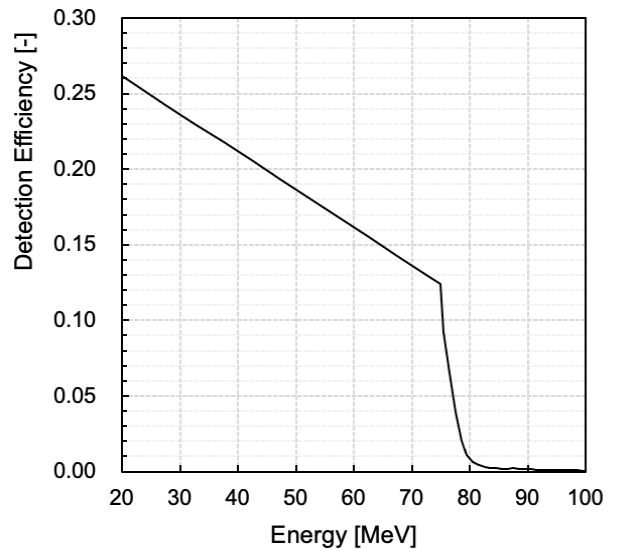

**Figure 5.** Detection efficiency curve of the normal mode of FAMES. The clear cutoff energy, 75 MeV, is determined by the minimum energy to penetrate Center PS, which is 20-cm high.

*3.3 ΔE–E analysis for random coincidence events*

As mentioned in Section 3, the measurement results of this study include random coincidence of electrical noises and CR electrons and positrons.

The tail of the electron and positron distribution slightly overlaps with muon events, but this overlap is sufficiently small to be ignored.

In addition, electrical noise of FAMES will appear around the low-energy region, both for Top and Center PSs. The noisy region will be disposed in the data analysis. Details are presented in Section 5.

*3.4 Normal mode analysis*

Normal mode data analysis is performed by dividing the measured deposit energy spectrum by a detection efficiency function, below 75 MeV. In this mode, all muons that have less than 75 MeV of kinetic energy can deposit all the energy if they travel toward the bottom surface of Center PS. Deposited energy can be underestimated only when a muon escapes from the side surface Center PS, but such events are removed by veto signals generated by Bottom PS. Thus, the recorded deposited energy can be considered equal to the incident energy. The escape ratio was predicted by PHITS using mono energy muons from 20 to 100 MeV. Figure 5 shows the efficiency curve derived from the escape ratio. As mentioned in Subsection 2.2, there is a clear cutoff energy, 75 MeV, as determined by the anticoincidence of the Bottom PS signal.

*3.5 Degrading mode analysis*

Data analysis of the degrading mode requires an inverse problem analysis to derive the energy differential muon flux $\phi(E)$, expressed as follows:

$$D(E_d) = \int R(E)\phi(E)\mathrm{d}E, \qquad (1)$$

where $D(E_d)$ and $R(E)$ represent the deposited energy spectrum and detection response function, respectively. We derived the response function using PHITS for two lead-degrader thicknesses, 5 and 20 cm (Figure 6). Notably, events produced by CR electrons and positrons are negligible because the 5-cm-thick lead can absorb 99% of CR electrons and positrons.

We used the FORIST code [20] for the inverse problem analysis procedures. FORIST has a long history, since 1976, of estimating fast neutron energy spectra from the pulse height distribution measured using a liquid scintillator (frequently, such inverse problem analysis is called "unfolding process" in neutron measurement field). FORIST uses the least-squares method, which does not require any initial guess solution in contrast to other unfolding algorithms. However, the solution is highly sensitive to fluctuations in the measured spectrum by noise and statistical uncertainty. Further, the energy range of response functions is critical because higher energy particles also induce drastic reactions to lower energy ranges. Thus, the measured spectrum is processed by the Gaussian smoothing method with a resolution parameter, $\sigma_k$, given by

$$\sigma_k = \frac{w_k E_k}{235.5} \quad (2)$$

where $w_k$ and $E_k$, respectively, denote a window width in percent (user parameter) and particle energy, i.e., the window width is the same as the full-width-at-half-maximum the smoothing function.

We chose 50% for the window width to obtain robust solution by grid search between 10 and 100%. The maximum and minimum energies for response functions were determined based on the following steps.

1) Derive the $E$-MeV muon detection probability $p(E)$ as the normalized response function:

$$p_{\text{res}}(E) = \frac{R(E)}{\int_0^\infty R(E)dE} \quad (3)$$

2) Calculate the cumulative probability distribution $P(E)$ as,

$$P(E) = \int_0^E p_{\text{res}}(E')dE'. \quad (4)$$

3) The energy range in the three-sigma confidence interval was considered in the unfolding process. The minimum and maximum energy, $E_{min}$ and $E_{max}$ are set to the energy that $P(E)$ returns 0.003 and 0.997 respectively.

In summary, when using the 5- and 20-cm-thick lead blocks, the energy ranges of response are 100–180 and 300–400 MeV, respectively.

## 4. Experiment

The low-energy muon flux measurement was performed at a five-story building (D-building), the Chikushi campus of Kyushu University, Fukuoka, Japan. The building is located at 130.5° east longitude, 33.5° north latitude, and 39 m above sea level. The building is made of concrete whose thickness is 20 and 15 cm for the floor and wall, respectively. FAMES was placed 200 cm away from any of the building's walls so as not to be affected for the detection acceptance.

As mentioned above, because we aimed to obtain a muon spectrum up to 400 MeV, the degrading mode measurements are necessary in addition to the normal mode. The appropriate degrader thickness is 20 cm in case the absorbers are made of lead. Moreover, the 5-cm-thick degrader condition was also performed to check whether resultant fluxes with the normal mode and the other mode are smoothly connected. Table 1 summarizes the experimental conditions.

**Table 1.** FAMES has normal (no degrader) and degrading modes (two conditions, 5- and 20-cm-thick lead blocks). The dynamic energy range for each mode is listed

| Mode | Degrader thickness (cm) | Measurable energy (MeV) | Measurement time (day) |
|---|---|---|---|
| Normal | – | ~75 | 10 |
| Degrading | 5 | 100–180 | 11 |
|  | 20 | 300–400 | 11 |

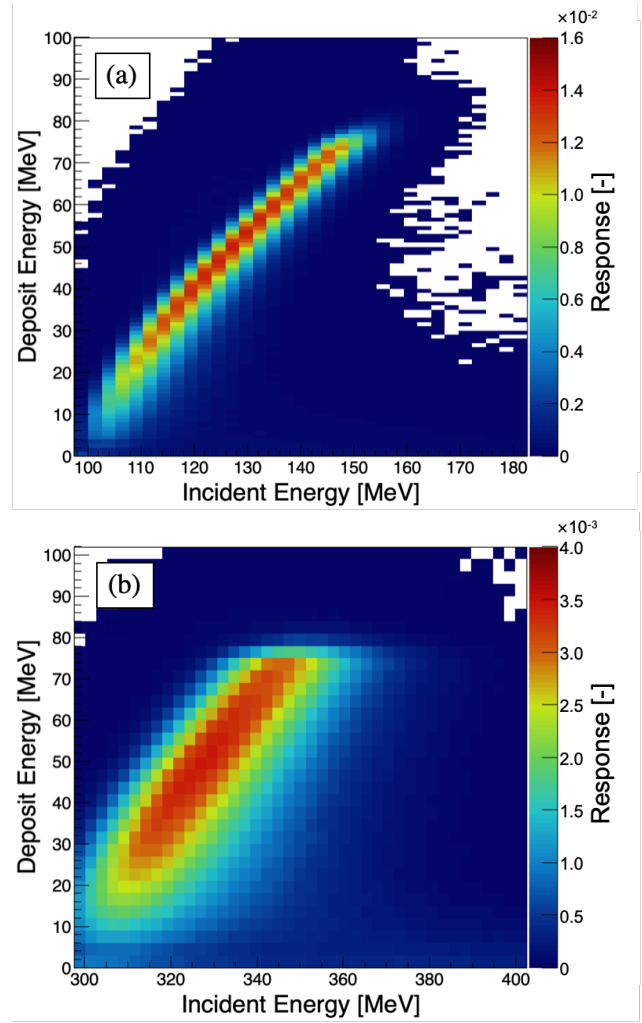

**Figure 6.** Two dimensional response functions of the degradation calculated using PHITS for (a) 5- and (b) 20-cm lead degraders. Owing to Coulomb scattering, the 20-cm lead response has a wider region of full-absorption events.

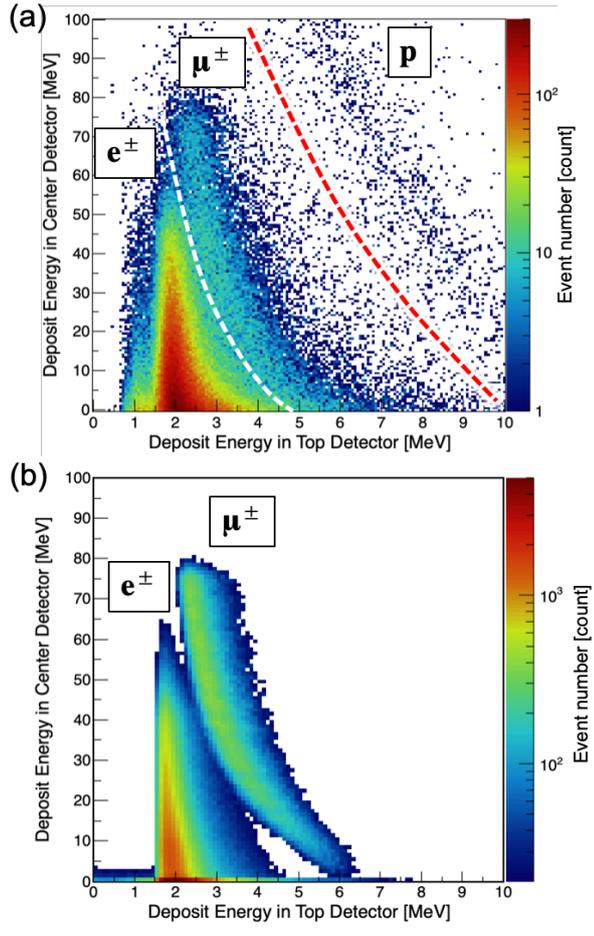

**Figure 7.** (a) Two dimensional histogram for Top and Center PSs to distinguish CR muons from CR electrons, positrons, and protons using the ΔE–E method. (b) Simulated ΔE–E distribution of CR muon, CR electrons and positrons. If other events are observed in the low-energy region, they can be regarded as random coincidences of electrical noises.

## 5 Results and discussion

### 5.1 ΔE–E plot for event identification

Figure 7 (a) shows the experimental ΔE–E plot, that has three components, CR electrons and positrons, muons, and CR protons. The CR protons are ignorable because they never overlapped with CR muon events. Thus, we performed PHITS simulation to obtain a ΔE–E plot only for CR muons and CR electrons and positorons (see Figure 7 (b)) whose double-differential fluxes were predicted by PARMA. The distribution of events by measurement is similar to that of simulation. We used all events above 20 MeV because they are well-separated from each other. In addition, the region was never smeared by the electrical noise events. The banana cut was adopted for the evemt separation.

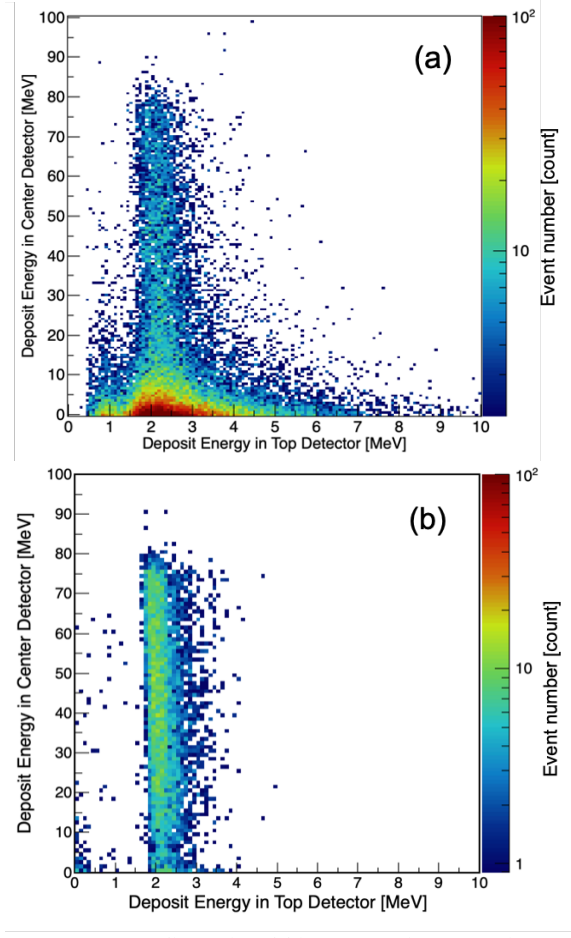

**Figure 8.** Two dimensional histogram of (a) experiment and (b) simulated data for Top and Center PSs using the 5-cm lead degrader. Almost all cosmic-ray electrons and positrons are absorbed in the degrader.

Contrary to the normal mode, the degrading modes are slightly affected by electrons and positrons in the ΔE–E plots, as shown in Figures 8 and 9 for the 5- and 20-cm-thick degraders, respectively.

### 5.2 Discussion

The resultant fluxes for all measruemnts are shown in Figure 10 with prediction by PARMA model. The PARMA estimation is similar to the present measured spectrum both in magnitude and the trend. For detailed discussion, we separated the measured energy range into four groups. Then, to calculate C/E, integral fluxes over each energy range for PARMA prediction and experimental spectrum (see Table 2).

The C/E values for Group 2 and 3 are almost 1 within the margin of their statistical uncertainty. In contrast, the integral values of the PARMA prediction have statistical significance from the present experimental data, underestimation for Group 1 and overestimation for Group 4. The underestimation range, i.e., the lowest part, is especially notable because it guides underestimation in prediction of soft error rate.

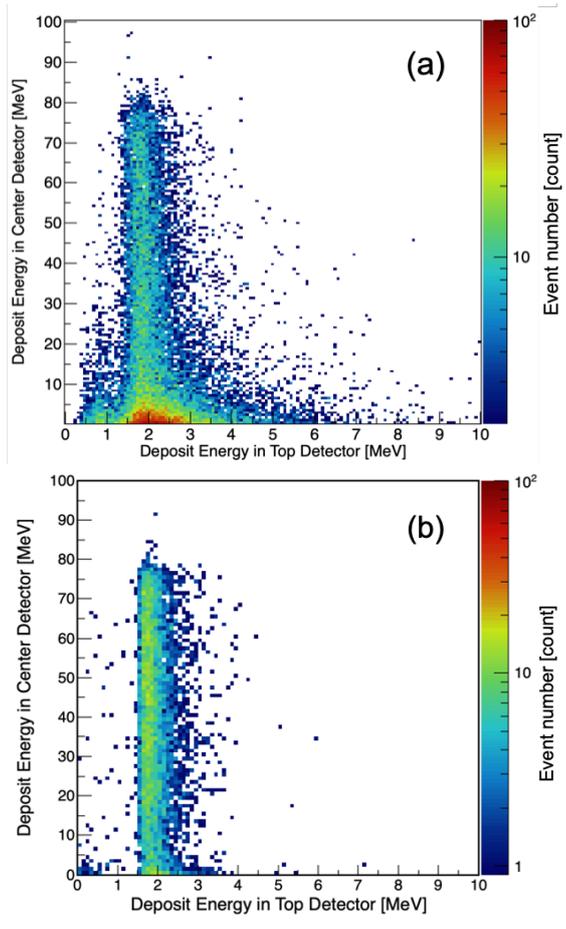

**Figure 9.** Two dimensional histogram of (a) experiment and (b) simulated data for Top and Center PSs using the 20-cm lead degrader.

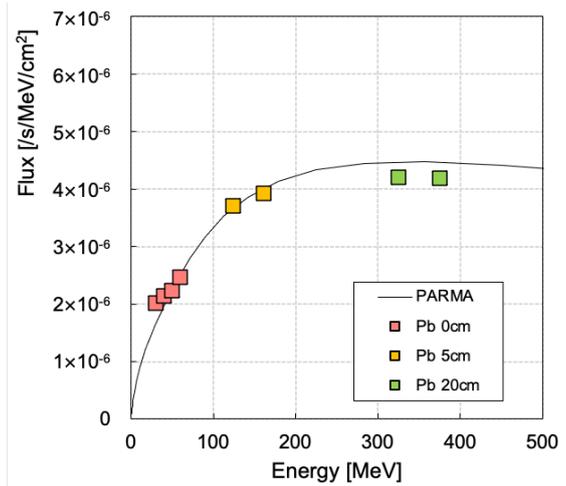

**Figure 10.** Comparison of measured results with the analytical model, PARMA. Red circles represent measurements without lead, green circles represent measurements with 5-cm-thick lead, blue circles represent measurements with 20-cm-thick lead, and the solid black line represents PARMA.

**Table 2.** Comparison of fluxes of CR muons obtained by PARMA model and experiment.

| Group ID | Energy (MeV) | Mode | Statistical uncertainty | C/E |
|---|---|---|---|---|
| 1 | 20–40 | Normal | 3-4% | 0.869 |
| 2 | 40–75 | Normal | 2-3% | 0.987 |
| 3 | 100–180 | Degrading (5 cm) | ~2% | 1.010 |
| 4 | 300–400 | Degrading (20 cm) | ~2% | 1.071 |

## 6 Conclusion

FAMES was developed to measure terrestrial cosmic-ray (CR) muons below a few hundreds of MeV, as required in soft error and CR muons studies. FAMES comprises three PSs to select the full-energy absorption events due to CR muons. We measured the energy differential CR muons spectrum ranging from 20 to 400 MeV. PARMA model prediction fairly agree with the present experimental spectrum, but underestimation takes place around 20-40 MeV range. Since such low energy muons have potential to induce soft error, the underestimation can also be a cause of underestimation of soft errors.

In the future, we plan to measure the double-differential spectra of muons by incorporating a muon tracking detector. For instance, adding two position-sensitive detectors is a simple but promising technique. We have developed a detection system named Cosmic Bench, which we are still experimenting with; the results to be obtained will be compared with results by FAMES to evaluate the analytical cosmic-ray flux model.

## Acknowledgements

We would like to thank Prof. Yukinobu Watanabe for useful discussions. We are greatful to Dr. Shouhei Araki for assistance with unfolding analysis.